\documentclass[runningheads]{llncs}

\usepackage[a4paper,hmargin=3.2cm,vmargin=3.5cm]{geometry}

\usepackage[T1]{fontenc}
\usepackage{graphicx}
\usepackage{amsmath}
\usepackage{amssymb}
\usepackage{booktabs}
\usepackage{array}
\usepackage[caption=false]{subfig}

\graphicspath{{figures/}}

\begin{document}

\title{A Multi-Stage Linear Programming Framework for Three-Phase State Estimation in Low-Voltage Distribution Grids \thanks{The paper was presented at the International Energy Future Conference, September 17–19, 2026, Rome, Italy}
} 
\titlerunning{Multi-Stage LP for Three-Phase LV State Estimation}

\author{Mohsen Banaei$^{*}$\and Eghbal Hosseini $^{*}$\and Ahlam Jameel $^{*}$\and Konstantin Filonenko $^{*}$\and Shahrzad Pour $^{*}$\and Razgar Ebrahimy $^{*}$}
\authorrunning{M. Banaei et al.}
\institute{Technical University of Denmark, Department of Applied Mathematics and Computer Science, Copenhagen, Denmark \\
\email{moban@dtu.dk}}

\maketitle

\begin{abstract}
Low-voltage (LV) distribution feeders are increasingly difficult to monitor because real-time load data are unavailable, historical measurements are sparsely sampled, and high-rate voltage sensors cover only a few nodes. This paper proposes a multi-stage linear programming (MSLP) estimator for three-phase unbalanced LV grids that reconstructs per-phase nodal voltages under such limited observability. The estimator linearises the three-phase power flow through voltage-to-power sensitivity matrices and adjusts the nodal active and reactive power injections so that the resulting voltages match the available measurements. To limit the error introduced by linearisation, the sensitivities are rebuilt and the power corrections refined over successive iterations, and power-balance constraints at intermediate metered nodes are added to tighten the feasible region. The method is validated on a real 50-node Danish LV feeder, where it attains a mean absolute error of 0.641~V, a standard deviation of 0.816~V, a root-mean-square error of 0.842~V, and a maximum error of 3.437~V. Compared with a single-stage linear estimator, MSLP lowers the error metrics by roughly 16\% on average, and additional studies quantify the influence of the input-preparation strategy and of the number of estimation meters.
\end{abstract}

\section{Introduction}
LV distribution feeders are absorbing a growing share of rooftop photovoltaics, heat pumps, and electric-vehicle chargers, whose bidirectional and volatile flows push nodal voltages toward statutory limits and make the operating state hard to anticipate~\cite{madsen2025demand,blomgren2023intensive}. Keeping such feeders safe requires continuous knowledge of the nodal voltages, yet LV grids remain sparsely instrumented: real-time load measurements are seldom available, historical consumption is logged only as coarse hourly aggregates, and high-rate voltage sensors cover a handful of nodes at best~\cite{dehghanpour2019survey,primadianto2017review}. Recovering the full set of nodal voltages from this sparse and delayed information is the task of distribution system state estimation (DSSE). Classical weighted-least-squares and Kalman-filter estimators were devised for well-measured transmission networks and lose accuracy in observability-poor LV settings~\cite{zhao2019robust,xu2018fast}. Optimisation-based estimators have therefore gained ground: linear and convex formulations scale to large feeders and converge deterministically~\cite{shaghaghi2024state,yadav2023review}, whereas learning-based estimators built on deep or graph neural networks can be accurate but need large training sets and may generalise poorly when the operating point drifts~\cite{kim2022distribution,menke2019distribution,cao2020fast,shamseldein2026hybrid,moshtagh2025topology,bhusal2021deep}. Metaheuristic and hybrid solvers have also been explored for the nonlinear power-flow equations~\cite{hassanein2021estimation,adewuyi2022power}. A recurring difficulty is the explicit treatment of three-phase unbalance under tight runtime and data-availability constraints.

This paper builds on our single-stage linear programming estimator~\cite{banaei2026linear} and our hybrid-metaheuristic estimator~\cite{banaei2026threestage}, and addresses the main weakness of the single-stage linear approach: the sensitivity coefficients that linearise the three-phase power flow are valid only for small excursions around the linearisation point, so a single linear approach accumulates error when the required power corrections are large. We propose MSLP estimator that re-linearises the power flow around the updated operating point at every iteration and drives the voltage mismatch down progressively. The main contributions are: (i) an iterative, sensitivity-based linear programming estimator that repeatedly rebuilds the voltage-to-power sensitivity matrices and refines the power corrections, keeping every subproblem linear while limiting linearisation error; (ii) the incorporation of active- and reactive-power balance constraints at intermediate metered nodes, which exploit the through-power recorded by non-terminal meters to shrink the feasible region; and (iii) an evaluation on a real 50-node Danish LV feeder that quantifies the accuracy gain over the single-stage estimator and studies the effect of the input-preparation strategy and of the number of estimation meters.

\section{System Model and Problem Definition}
\label{sec:system}
\begin{figure}[!htb]
\centering
\includegraphics[width=0.62\textwidth]{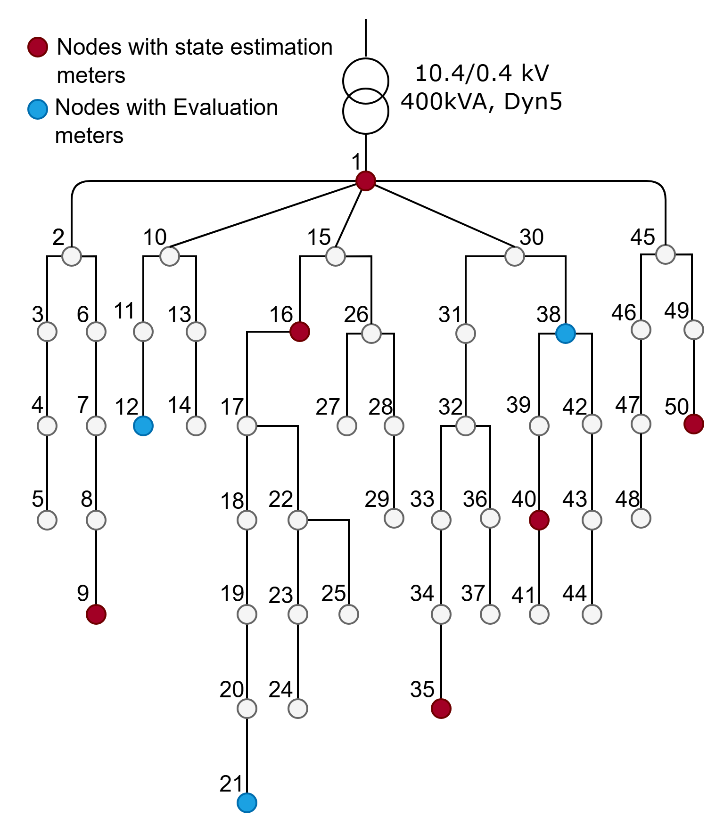}
\caption{Single-line diagram of the studied 50-node three-phase LV feeder, showing the placement of state-estimation and evaluation meters.}
\label{fig:grid}
\end{figure}

The method is developed for the three-phase unbalanced LV feeder used throughout the SmartDSO project, shown in Fig.~\ref{fig:grid}. The network comprises 50 nodes arranged as radial branches fed from the secondary of a 400~kVA, 10.4/0.4~kV Dyn5 distribution transformer; each node aggregates the load of a small group of buildings behind a fuse box. Consistent with real distribution-system-operator practice, only a few nodes are metered: six nodes $\{1,9,16,35,40,50\}$ carry state-estimation meters whose per-phase voltages feed the estimator, and three further nodes $\{12,21,38\}$ carry evaluation meters that are withheld from the estimation and used only to score accuracy. Because node~16 lies inside the feeder rather than at a terminal, the power flowing through its meter bounds the consumption of all nodes supplied through it, a property exploited later by the balance constraints of Section~\ref{sec:method}.

The estimator requires per-phase active and reactive power injections at every node, which are not directly measured. Following the project pipeline~\cite{banaei2026linear,banaei2026threestage}, these are obtained in two steps. First, the aggregate active and reactive demand at each node is inferred from historical smart-meter data, either by short-term forecasting~\cite{hosseini2025optimized,hosseini2025optimal} or, as studied in Section~\ref{sec:results}, by averaging over matching hours and weekdays. Second, each nodal quantity is disaggregated across phases L1--L3 using phase-share coefficients estimated at metered buses and propagated to unmetered buses through impedance-based clustering. Given these per-phase injections and the sparse real-time voltage measurements, the objective is to reconstruct the per-phase voltage magnitude at every node. The three-phase power flow that links injections to voltages is nonlinear and nonconvex, which is the central obstacle addressed by the proposed method.

\section{Proposed Multi-Stage Linear Programming Method}
\label{sec:method}
In this section, the proposed method for three-phase grid state estimation is presented. The core idea is to modify some parameters of the grid so that the power-flow result of the modified grid becomes as close as possible to the voltage measurements at the nodes equipped with state-estimation meters. Here, the active and reactive power consumptions of the nodes are taken as the decision variables. Modifying these parameters optimally requires the grid equations; however, the three-phase grid equations are non-linear and non-convex. To overcome this issue, two matrices are constructed that estimate the relation between each node's active or reactive power consumption and the voltage at each node equipped with a state-estimation meter. The following steps are used to calculate these matrices:
\begin{enumerate}
\item First, the grid topology information is used to create a grid model in OpenDSS, an open-source software for modelling distribution grids that provides an interface with Python.
\item Then, in Python, at each time interval the estimated active and reactive powers obtained in the disaggregation stage are fed to the model, and a baseline for the nodal per-phase voltages $V_{i,t}^{base}$ is calculated by running the power flow in OpenDSS.
\item Next, the active (reactive) power of each node $j\in N$ ($N$ is the set of all phases of all nodes) is increased by $\partial P$~kW ($\partial Q$~kVar), and its impact on the voltage of each node with a meter $i\in N_{m}$ ($N_{m}$ is the set of all phases of all nodes with meters) is recorded as $a_{i,j}$ ($b_{i,j}$):
\begin{equation}
a_{i,j}=\frac{\partial V_{i}}{\partial P_{j}}, \qquad b_{i,j}=\frac{\partial V_{i}}{\partial Q_{j}} .
\end{equation}
\end{enumerate}
Using these matrices, the estimated voltage can be formulated as:
\begin{equation}
V_{i}^{est}=V_{i}^{base}+\sum_{j\in N}a_{i,j}\,\Delta P_{j}+\sum_{j\in N}b_{i,j}\,\Delta Q_{j}.
\end{equation}

The key point about this linearisation is that it is valid only for small changes in $\Delta P_{j}$ and $\Delta Q_{j}$; as these values increase, the estimation error increases as well. The proposed solution is to limit the variations of $\Delta P_{j}$ and $\Delta Q_{j}$ within the optimisation problem and to update the components of the matrices across iterations. At each iteration $n$, the following optimisation problem is solved to minimise the difference between the measured and estimated voltages and to obtain the optimal values of $\Delta P_{j}^{n}$ and $\Delta Q_{j}^{n}$:
{\small
\begin{align}
\min\quad & A\sum_{j\in N}\!\left(\Delta P_{j}^{abs,n}+\Delta Q_{j}^{abs,n}\right)+B\left(\Delta P^{\max}+\Delta Q^{\max}\right)+\sum_{i\in N_{m}}X_{i}\\
\text{s.t.}\quad & V_{i}^{est}=V_{i}^{base,n}+\sum_{j\in N}a_{i,j}^{n}\Delta P_{j}^{n}+\sum_{j\in N}b_{i,j}^{n}\Delta Q_{j}^{n}, & \forall\, i\in N_{m}\\
& -X_{i}\le V_{i}^{est}-V_{i}^{meter}\le X_{i}, & \forall\, i\in N_{m}\\
& -\Delta P_{j}^{abs,n}\le \Delta P_{j}^{n}\le \Delta P_{j}^{abs,n}, & \forall\, j\in N\\
& -\Delta Q_{j}^{abs,n}\le \Delta Q_{j}^{n}\le \Delta Q_{j}^{abs,n}, & \forall\, j\in N\\
& \sum_{j\in N_{\phi,k}^{mid}}\!\left(P_{j}^{base,n}+\Delta P_{j}^{n}\right)\le P_{\phi,k}^{meter}, & \forall\, k\in N_{\phi}^{mid}
\end{align}
\begin{align}
& \sum_{j\in N_{\phi,k}^{mid}}\!\left(Q_{j}^{base,n}+\Delta Q_{j}^{n}\right)\le Q_{\phi,k}^{meter}, & \forall\, k\in N_{\phi}^{mid}\\
& \Delta P_{j}^{abs,n}\le \Delta P^{\max}, & \forall\, j\in N\\
& \Delta Q_{j}^{abs,n}\le \Delta Q^{\max}, & \forall\, j\in N\\
& \Delta P^{\max}\le k, & \\
& \Delta Q^{\max}\le k, &
\end{align}
}%
where $\Delta P_{j}^{n}$ and $\Delta Q_{j}^{n}$ represent the changes in the estimated active and reactive powers at iteration $n$, $\Delta P_{j}^{abs,n}$ and $\Delta Q_{j}^{abs,n}$ are their absolute values, and $\Delta P^{\max}$ and $\Delta Q^{\max}$ are the maximum changes in active and reactive power over all nodes; the balance constraints (9)--(10) hold for each phase $\phi\in\{1,2,3\}$. $X_{i}$ is a penalty for the cases where the power changes cannot reduce the voltage difference to zero.

Objective function (4) minimises the error between the estimated and metered voltages together with the changes in active and reactive powers, which improves the performance of the estimator. Constraint (6) calculates the error between the estimated and metered voltages. Constraints (7) and (8) obtain the absolute values of $\Delta P_{j}^{n}$ and $\Delta Q_{j}^{n}$, respectively. Constraints (9) and (10) satisfy the power balance at the metered nodes: for a node with a meter that is not located at the grid edge, such as node~16, the power that goes through the meter is available, and this (active or reactive) power must be greater than the sum of the (active or reactive) power consumption of the nodes supplied through it (nodes 17--25). The set of nodes connected to phase $\phi$ of a middle metered node $k$ is called $N_{\phi,k}^{mid}$, and $N_{\phi}^{mid}$ represents the set of all middle nodes. Constraints (11) and (12) determine the maximum values of $\Delta P_{j}^{n}$ and $\Delta Q_{j}^{n}$ that are minimised in (4) to improve the performance of the method. Constraints (13) and (14) limit the maximum changes in active and reactive power to $k$; this parameter keeps the range of $\Delta P_{j}^{n}$ and $\Delta Q_{j}^{n}$ within values that do not cause significant errors in computing $V_{i}^{est}$.

\begin{figure}[!htb]
\centering
\includegraphics[width=0.50\textwidth]{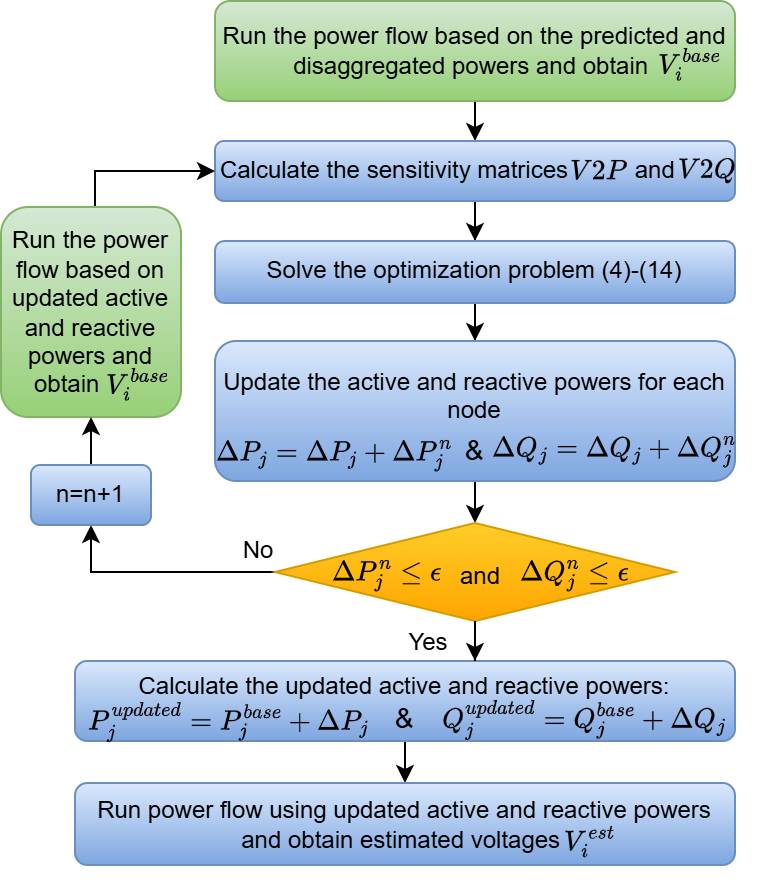}
\caption{Iterative workflow of the proposed multi-stage linear programming estimator.}
\label{fig:flow}
\end{figure}

The iterative methodology is illustrated in Fig.~\ref{fig:flow}. 
At each iteration, the matrix elements $a_{i,j}$ and $b_{i,j}$ are computed from the modified grid state, the optimisation problem (4)--(14) is solved, and $\Delta P_{j}$ and $\Delta Q_{j}$ for each node are updated as
\begin{equation}
\Delta P_{j}\leftarrow \Delta P_{j}+\Delta P_{j}^{n},\qquad \Delta Q_{j}\leftarrow \Delta Q_{j}+\Delta Q_{j}^{n}.
\end{equation}
If the changes in these parameters are not less than a threshold, a power flow is run using the updated values, the new state $V_{i}^{base}$ is calculated, and the process is repeated. Otherwise, the final modified active and reactive powers are calculated, a power flow is run, and the resulting voltages are taken as the estimated per-phase nodal voltages.

\section{Results and Discussion}
\label{sec:results}
The proposed estimator is assessed by comparing the estimated voltages at the evaluation nodes against the corresponding meter readings. To cover a range of operating conditions, the estimation is carried out for one representative day, 1~October~2025, at 15-minute resolution. Node~16 is a middle node, so constraints (9)--(10) are active for it.

\subsection{State Estimation Accuracy}
\begin{figure}[!htb]
\centering
\includegraphics[width=0.95\textwidth]{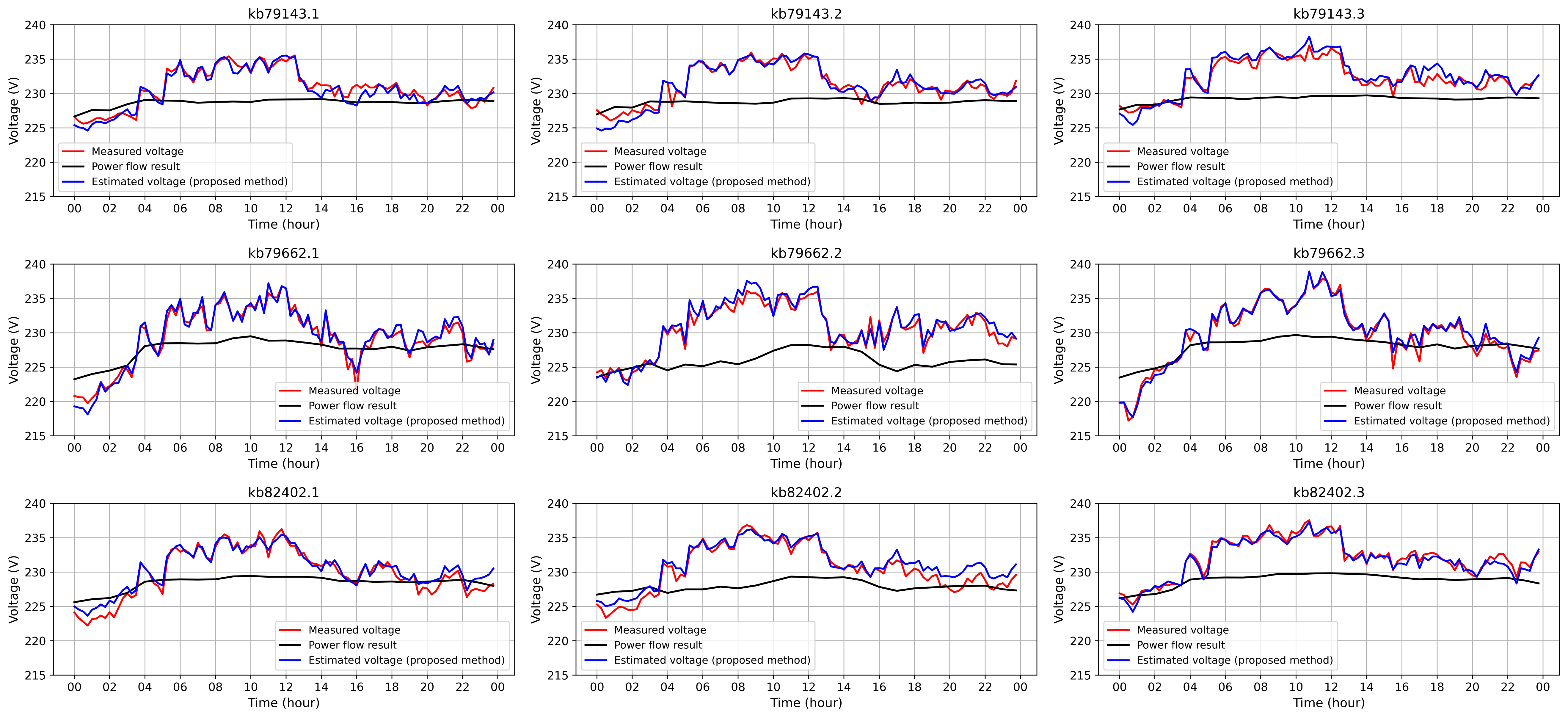}
\caption{Measured voltage, power-flow result, and estimated voltage at the three evaluation meters. Rows correspond to evaluation nodes 12, 21, and 38 (top to bottom) and columns to phases L1, L2, and L3 (left to right).}
\label{fig:volt}
\end{figure}

\begin{figure}[!htb]
\centering
\includegraphics[width=0.45\textwidth]{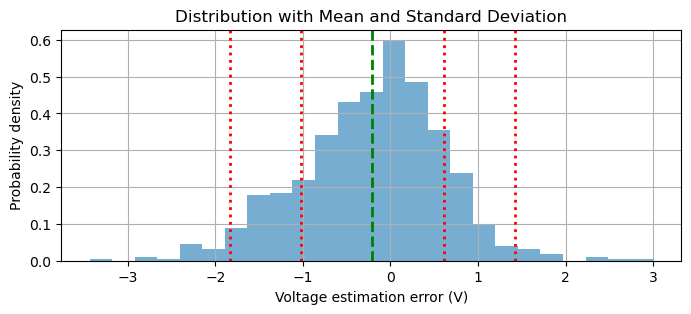}
\caption{Distribution of the voltage estimation error with its mean (green) and standard deviation (red).}
\label{fig:hist}
\end{figure}

Figure~\ref{fig:volt} compares, for the three phases of each evaluation meter, the measured voltage, the power-flow result obtained from the forecast injections, and the output of the proposed estimator. While the raw power-flow curve fails to follow the realised voltages, the estimated profile tracks the measurements closely across all phases. Four metrics are used to quantify accuracy: mean absolute error (MAE), standard deviation (STD), root-mean-square error (RMSE), and maximum error (ME). For the studied case the estimator attains an MAE of 0.641~V, an STD of 0.816~V, an RMSE of 0.842~V, and an ME of 3.437~V; in 96\% of the samples the error stays below 1.632~V (0.7\%). The error distribution together with its mean and standard deviation is shown in Fig.~\ref{fig:hist}.


\subsection{Comparison with the Single-Stage Estimator}
The proposed multi-stage estimator (MSLP) is compared with the single-stage linear programming estimator (LP) of~\cite{banaei2026linear} ~\cite{team} over four dates chosen according to data availability. A mean runtime (MRT) per solve is added to the four accuracy metrics. As summarised in Table~\ref{tab:compare}, MSLP improves the MAE, STD, RMSE, and ME on average by about 16.55\%, 17.31\%, 16.00\%, and 9.81\%, respectively. This accuracy gain comes at a computational cost: the iterative refinement makes MSLP roughly 2.5 times slower than the single-stage method.

\begin{table}[!htb]
\centering
\caption{Accuracy and runtime of the proposed MSLP method versus the single-stage LP method over four dates.}
\label{tab:compare}
\resizebox{\textwidth}{!}{%
\begin{tabular}{l cc cc cc cc c}
\toprule
 & \multicolumn{2}{c}{Oct 1, 2025} & \multicolumn{2}{c}{Nov 19, 2025} & \multicolumn{2}{c}{Dec 15, 2025} & \multicolumn{2}{c}{Jan 7, 2026} & Avg.\\
\cmidrule(lr){2-3}\cmidrule(lr){4-5}\cmidrule(lr){6-7}\cmidrule(lr){8-9}
Metric & MSLP & LP & MSLP & LP & MSLP & LP & MSLP & LP & impr.\\
\midrule
MAE (V)   & 0.641 & 0.785 & 0.812 & 1.006 & 0.807 & 0.944 & 0.867 & 1.009 & 16.55\%\\
STD (V)   & 0.816 & 0.999 & 1.105 & 1.389 & 1.052 & 1.241 & 1.129 & 1.332 & 17.31\%\\
RMSE (V)  & 0.842 & 0.999 & 1.135 & 1.377 & 1.056 & 1.259 & 1.141 & 1.336 & 16.00\%\\
ME (V)    & 3.437 & 3.853 & 7.581 & 7.116 & 4.444 & 4.875 & 4.457 & 6.033 & 9.81\%\\
MRT (s)   & 8.662 & 3.014 & 9.612 & 3.307 & 8.226 & 5.548 & 8.794 & 3.270 & $-148.82\%$\\
\bottomrule
\end{tabular}%
}
\end{table}

\subsection{Effect of the Input-Preparation Strategy}
Table~\ref{tab:input} reports the metrics when the real-time active and reactive powers are prepared by forecasting versus by averaging matching hours and weekdays over two months of history. For this feeder and choice of evaluation meters, averaging slightly outperforms forecasting. Since each node represents only a few buildings, their consumption is hard to forecast and the forecast error can exceed the error introduced by simple averaging. This outcome is specific to the studied grid and meter configuration and may differ for other feeders.

\begin{table}[!htb]
\centering
\caption{State estimation metrics with forecasting versus historical averaging for input preparation.}
\label{tab:input}
\begin{tabular}{lcc}
\toprule
Metric & Forecast & Averaging\\
\midrule
MAE (V)  & 0.641 & 0.613\\
STD (V)  & 0.816 & 0.799\\
RMSE (V) & 0.842 & 0.769\\
ME (V)   & 3.437 & 3.001\\
\bottomrule
\end{tabular}
\end{table}

\subsection{Effect of the Number of Estimation Meters}
Table~\ref{tab:meters} lists the metrics as estimation meters are progressively removed. As expected, fewer meters increase the estimation error; halving the number of estimation meters raises the metrics by about 68\% on average, which illustrates the sensitivity of the estimator to metering coverage.

\begin{table}[!htb]
\centering
\caption{State estimation metrics for a decreasing number of estimation meters.}
\label{tab:meters}
\begin{tabular}{lcccc}
\toprule
Case & MAE (V) & STD (V) & RMSE (V) & ME (V)\\
\midrule
Base case (6 meters)                 & 0.641 & 0.816 & 0.842 & 3.437\\
5 meters (remove 1)                  & 0.878 & 1.106 & 1.104 & 4.419\\
4 meters (remove 1, 9)               & 0.984 & 1.199 & 1.218 & 4.184\\
3 meters (remove 1, 9, 40)           & 1.090 & 1.338 & 1.400 & 5.845\\
\bottomrule
\end{tabular}
\end{table}

\section{Conclusion}
This paper presented a multi-stage linear programming framework for per-phase voltage estimation in three-phase unbalanced LV distribution grids under limited observability. By re-linearising the power flow around the updated operating point at each iteration and enforcing power-balance constraints at intermediate metered nodes, the estimator keeps every subproblem linear while progressively reducing the linearisation error. On a real 50-node Danish feeder it achieved an MAE of 0.641~V and improved the accuracy metrics over a single-stage linear estimator by roughly 16\% on average, at the cost of a longer runtime. Complementary studies showed that, for this feeder, historical averaging can be a competitive alternative to forecasting for input preparation, and that accuracy degrades markedly as metering coverage is reduced. Future work will focus on lowering the computational cost of the iterative scheme and on extending the evaluation to larger feeders and more diverse operating conditions.

\subsection*{Acknowledgments}
This research was supported by SmartDSO project, which is funded under EUDP programme. Project ID: 640241-521879

\end{document}